\documentclass{ijuc}
\usepackage[]{graphicx}

\begin{document}

\title{Bacterial self-organisation and computation}

\author{Martyn Amos \inst{1}\email{M.R.Amos@exeter.ac.uk}, David A. Hodgson \inst{2} and Alan Gibbons \inst{3}}

\institute{$^{1}$Department of Computer Science, University of Exeter \\ $^{2}$School of Biological Sciences, University of Warwick, Coventry CV4 7AL, UK \\ $^{3}$Department of Computer Science, King's College, London WC2R 2LS, UK}

\def\received{Submitted to the {\it International Journal of Unconventional Computing}}

\maketitle

\section{Introduction}

Researchers in the field of complex systems are often
concerned with the study of collections of motile elements, such
as fish, birds, insects and cells \cite{cdfstb01}. These organisms may exhibit
characteristic patterns of self-organized collective behaviour
such as shoaling \cite{parrish02}, flocking \cite{reynolds87}, swarming \cite{okubo86} and other types of aggregation \cite{parrham97}.
We are particularly interested in cells, as the ability to reliably
predict and control the dynamical behaviour of cellular colonies
has profound implications, both on the study of the biological
system \textit{per se} and also on the potential for their use in
a diverse range of computational and engineering applications.

When millions of cells act collectively as a group (as is the
case in any colony), a large variety of different
patterns arise as a result of how the individual cells respond to
the others in their neighbourhood and to the conditions in their
environments (which will be different for different individual
bacteria, even if they are relatively close to each other) 
\cite{bb91,bb95,sha98,sha97}. The physical appearance of these patterns is largely due to cells aggregating together in different ways at different positions. These
patterns are the emergent result of local interactions and
environmental conditions, and can be usefully viewed as being
\textit{programmed} by both the particular way the cells in
question interact, and by the particular way they
respond to environmental signals (and, thus, their genetic ``program"). Also, 
these patterns are widely
regarded as functional; for example, in stress conditions (such as
a toxic chemical in the environment), the pattern formed will be
ideal for protecting a maximal number of individuals (e.g. by
minimising the number of cells without immediate neighbours) while
promoting the colony's ``search" for a less harmful environment
(by, e.g., extending ``arms" of cells which explore the
environment). The development of {\it spatially-extended} functional patterns purely as a result of {\it local} interactions between individuals and with their {\it immediate} environment is a form of computation which is extremely common and useful in nature, but which is poorly understood, not well-explored and underexploited in computer science".

The main purpose of the article is to highlight {\it chemotaxis} (cellular movement) as a rich source of
potential engineering applications and computational models, highlighting current research and possible
future work. We first give a brief description of the biological mechanism, before describing recent work on modelling it {\it in silico}. We then propose a methodology for
extending existing models and their possible application as a fundamental tool in {\it engineering} cellular pattern formation. We discuss possible engineering applications of human-defined cell patterns, as well as the potential for using abstract models of chemotaxis for generalised computation, before concluding with a brief discussion of future challenges and opportunities in this field.

\section{Bacterial chemotaxis}

The ability to sense and respond to changes in the environment is a fundamental property of
living organisms. Because of the basic importance of these capabilities for survival, nature has evolved many mechanisms to facilitate reaction to environmental signals. Such behaviours are often in response to environmental stress (e.g., extremes of temperature or light, changes in pH or the presence of toxic chemicals), and may include changes in gene expression \cite{gasch00}, immune response \cite{padgett03} or cellular movement \cite{bj00,bb91,tsimring95}. It is the last response that particularly interests us in this article.

\subsection{Chemotaxis mechanism}

{\it Chemotaxis} is the movement of motile cells \cite{berg93,berg00,mittal03} in response to concentrations of certain chemicals (see \cite{berg03} for a general overview).
The derivation of the {\it chemo} component of the term is obvious; {\it taxis} refers to {\it directed movement}, as opposed to chemo{\it kinesis}, which refers to the {\it rate} of movement.
 By ``motile", we mean a cell that is capable of {\it autonomous} movement; there are many
mechanisms by which a cell may move itself, including ``crawling" by shape deformation, the to-and-fro motion of hair-like structures on the cell surface, or by the whipping of a molecular ``propellor" (for a comprehensive overview of motility mechanisms, see \cite{bray01}).
This last method is used by the {\it E. coli} bacterium, as it possesses a number of thread-like {\it flagella} on its cell surface, which may rotate either clockwise or anti-clockwise under the probabilistic control of the cell. When individual flagella are rotated anti-clockwise, they form a single, coherent ``bundle", the movement of which propels the bacterium forwards. A bacterial ``run" is terminated when the flagellar bundle disintegrates as a result of its rotation being reversed. In this situation, the bacterium {\it tumbles} for a short time, a situation defined as ``erratic motion without net transition" \cite{mittal03}. When resuming directed movement after a tumble, the bacterium moves off in a new, effectively random direction. When on a run, the bacterium can therefore implement a biased random walk by manipulating the probability of either remaining running by continuing to rotate its flagella in the anti-clockwise direction, or of terminating the run by reversing the direction, causing it to tumble. In this way, a cell may explore its environment, making biased decisions about whether or not to continue moving in the current direction. In the absence of any chemical signals, individual bacteria perform a purely random walk \cite{mittal03}. However, a cell's choices may be
influenced by {\it chemotactants}; chemicals in the environment which may either {\it attract} or {\it repel} the bacterium. Bacteria respond to a chemotactant by sensing the chemical's concentration gradient and then reacting appropriately (for example, a bacterium may attempt to move up the gradient of an {\it attractant} by increasing the probability of running while the detected gradient increases). One immediate question that arises from this description is how a bacterium only 2-$\mu$m long can sense a chemical gradient with a decay distance of (perhaps) many millimetres. In 1972, Macnab and Koshland \cite{macnab72} proposed a bacterial ``temporal sensing mechanism" (i.e., a memory), which would allow bacteria to make comparisons between past and present environmental conditions. The existence of this mechanism was later confirmed by Segall, Block and Berg \cite{segall86}; their work demonstrated that {\it E. coli} possess a 4-second memory that, combined with single-molecule sensitivity, allows these bacteria to respond to environmental cues in an extremely sophisticated fashion.

\subsection{Chemotactic pattern formation}

\begin{figure}
\begin{center}
\scalebox{0.5}{\includegraphics{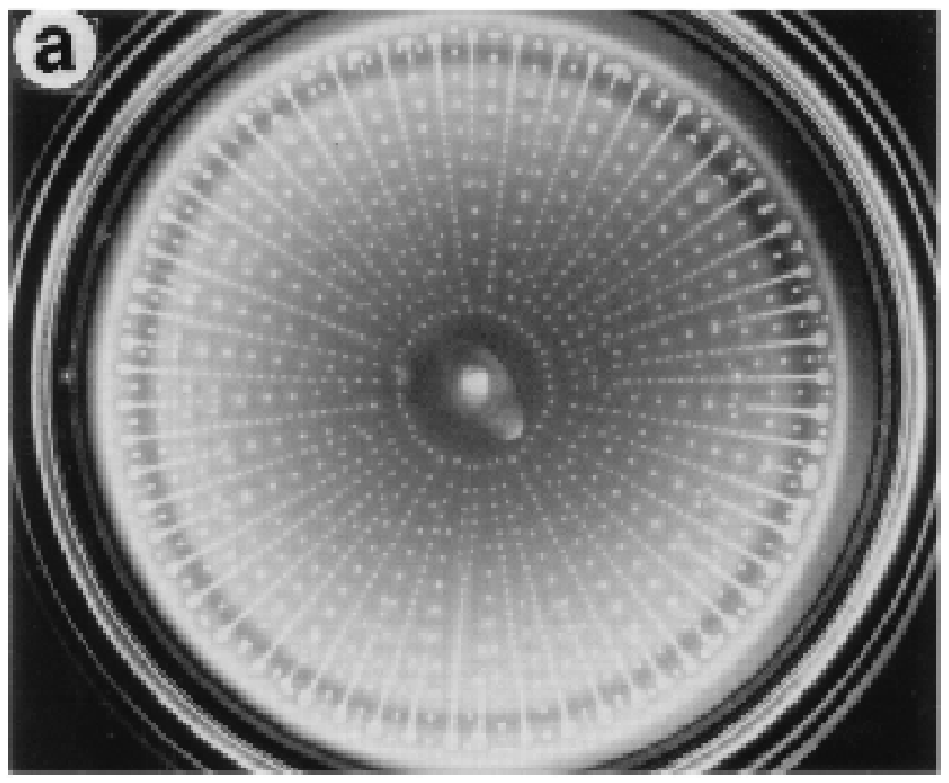}}\scalebox{0.5}{\includegraphics{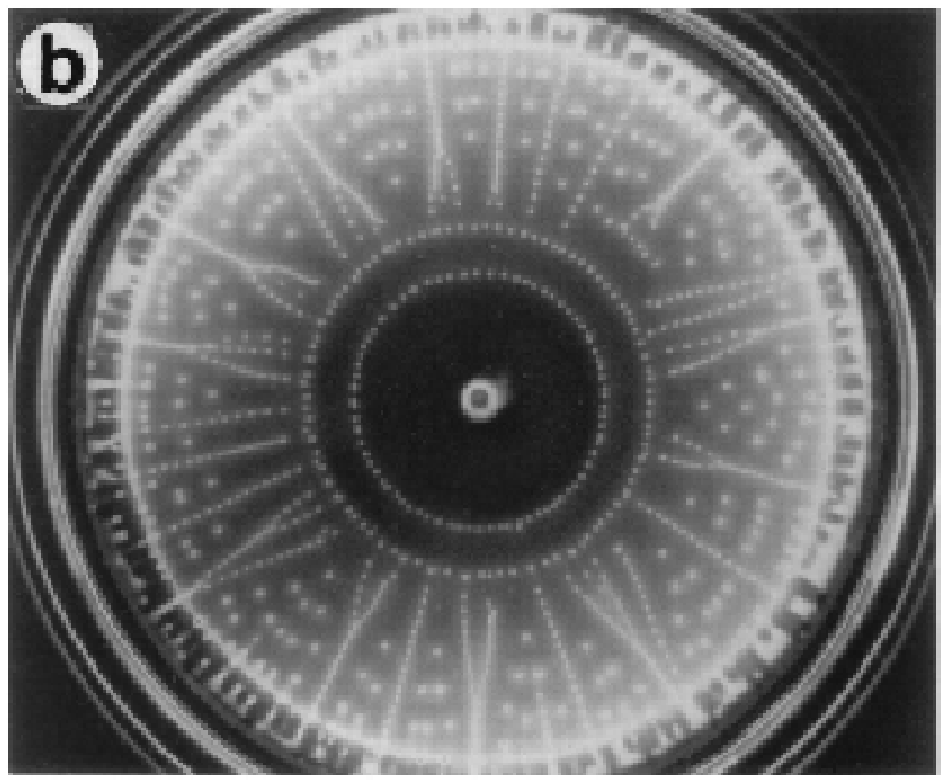}}
\end{center}
\caption{Differential spatial patterning of {\it Salmonella typhimurium} due to 0.2mM difference in environmental citrate concentration (taken from \cite{berg96}, and used with permission).}
\label{typh}
\end{figure}

For many years, it was believed that chemotaxis in motile bacteria served only to maximise their chances of encountering nutrients. Consider the situation where a colony is innoculated at a single central point in a dish of gel containing nutrients.  The colony will swarm outwards in either a single ring or a series of concentric rings \cite{bb91}, responding to spatial gradients of nutrient attractants. However, it is now clear that, under conditions of stress (e.g., exposure to hazardous byproducts of respiration, such as peroxides), {\it E. coli} themselves generate attractants,
becoming moving sources of chemical signals \cite{bb91,bb95,tsimring95}. This signal generation facilitates {\it inter-cellular communication}, leading to the collective formation of spatial patterns of spots, stripes, moving bands and other complex structures \cite{mittal03}. Example of such structures, generated by a cousin of {\it E. coli}, are depicted in Figure ~\ref{typh}.

The function of each type of pattern depends very much on the external pressure on the colony. Mittal {\it et al.} describe one scenario \cite{mittal03}, where {\it E. coli} form well-defined spotted clusters. Such patterns might emerge due to the presence of {\it xenobiotics} -- chemicals not usually found in the organism being studied (for example, drugs or carcinogenic compounds). In this situation, the bacteria may need to respond quickly by producing detoxifying enzymes \cite{brazil03}, high concentrations of which may only be produced in dense cell clusters. ``Clustering by chemotactic aggregation enables bacteria to
generate a robust spatial structure and maintain a localized population through the temporal
sensing of the concentration of a single species of molecule secreted by them..." \cite{mittal03}.

Ben-Jacob and Levine describe a different type of pattern formation in {\it Paenibacillus
dendritiformis}, bacteria that move over hard surfaces by excreting lubricant 
chemicals on which they ``skate" to move around \cite{bjlevine05}. A 
level of ``wetting" sufficient for collective movement requires the contributions of a large 
number of individuals. In situations where food is scarce, the required population density is
unsustainable, which creates a trade-off situation for the colony as a whole. This is resolved by  the population forming a branching structure; in each branch there are just enough bacteria to generate sufficient lubricant without exhausting the food supply. Ben-Jacob and Levine report
that individual bacteria constantly adjust their lubricant properties and production rate to
generate different branch structures, according to the environmental hardness and nutrient levels. They argue that this ``self-engineering" may be thought of as a form of collective constraint satisfaction, where the necessity of movement is balanced against the resources required for such migration.

\section{Modelling chemotaxis}

There exist several different \textit{in silico} approaches to studying collective cell behaviour. Many
previous studies model the cell population and its environment at a {\it macroscopic}
level, by describing the system using Partial Differential Equations. The classic model of this type is due to Keller and Segel \cite{keller71} (see \cite{horstmann03} for a comprehensive review). Numerical modelling approaches may, on the other hand, simulate chemotaxis at {\it either} the macroscopic population level, or the microscopic level of individual cells. Such models are generally based on finite difference/finite element methods (see \cite{hilpert05} for  a full list of relevant references). Lattice-Boltzmann models \cite{hilpert05} represent
bacteria and chemoattractants on a regular grid, and have been used with some success. However, such models fail to capture all aspects of real bacterial behaviour, and their application in unstructured (i.e., irregular) environments is still unclear. Finally, simulation studies that model individual cells (or groups of cells) have recently been developed \cite{bj94,emonet05,ginovart02,hoar03,sch02}.
It is this last approach that concerns us here, and we consider in particular the
{\it communicating walkers} model, first proposed by Ben-Jacob \textit{et al.} \cite{bj94}. The original model considers the following biological features: {\bf (1)} diffusion of nutrients,
{\bf (2)} bacterial movement, 
{\bf (3)} bacterial reproduction,
{\bf (4)} local communication.

Diffusion of nutrients is handled by solving
the diffusion equation for a particular nutrient concentration on
a triangular lattice. Bacteria are represented by ``walkers",
each of which is a coarse-grained sub-population of the colony.
Each walker is an {\it agent} \cite{wooldridge02} described by its location and an internal degree of
freedom (or ``internal energy"), which affects its activity.
Walkers lose energy at some constant rate, and may consume
nutrients where available. When the walker's internal energy
reaches zero it becomes stationary.
When food is sufficient, a walker's internal energy increases, and
when it reaches some threshold it divides (reproduction).
Stationary walkers produce a communication chemical at some fixed
rate (in an attempt to drive other bacteria away) and each of the
active walkers consumes the chemical at some fixed rate. The
movement of the active walkers is therefore a random walk with a
bias along the gradient of the communication field. Results
obtained using this simulation framework have been encouraging,
generating many of the patterning phenomena observed in the
laboratory \cite{bj98-ann,cohen98,cohen99}.

\subsection{Extensions to the model}

We now propose certain extensions to the basic communicating walkers model.
These are intended to improve the realism of the model, by incorporating
consideration of such factors as generation of nutrient
chemoattractant gradient by nutrient consumption, release and
consumption of chemoattractant, release and consumption of
chemorepellant, rheology of medium matrix and bacterial
reproduction. In detail, the additional features we propose
are {\bf (1)} Substrate structure. Standard agar gel is made up of
aggregates of helices that give it a porous structure, which may
produce effects that are unrelated to inter or intracellular processes.
The substitution of agar with materials such as carboxymethyl cellulose, removes this structure, giving a more homogenous environment and removing
the contribution of gel structure to pattern formation. The facility to vary this parameter would allow the possibility of studying
the effect of both gel strength and structure. {\bf (2)} Nutrient spectrum and
distribution. Previous work has assumed uniform nutrient distribution. It is clear that non-uniform distribution will
affect pattern formation in complex non-linear ways, so we must consider the possibility of ``doping" the media surface with regions of nutrient of
arbitrary complexity. {\bf (3)} Response to toxins. The Budrene and Berg patterns \cite{bb91,bb95}
have been shown to be due to the accumulation of respiration by-products such
as hydrogen peroxide when growing on organic acids as sole carbon and energy
source. The formation of micro-colonies aids in the more rapid destruction of
hydrogen peroxide by endogenous catalase. We can vary the speed of accumulation
and destruction of hydrogen peroxide by doping and incorporation of exogenous catalase in the medium, and examine the effect on pattern formation.
{\bf (4)} Genetic components. There are many mutants available in
\textit{E. coli}, that affect chemotaxis, nutrient use and toxin degradation.
The effect of these mutations on pattern formation can be assessed within the model
we propose, and would allow us to vary parameters as dictated by the discrete \textit{in silico}
modelling results.

\subsection{Benefits of this approach}

There are now many compelling reasons for considering this {\it individual-based} approach to the study of bacterial pattern formation. If we require additional biological realism,
adding extra components to a system of differential equations will
quickly render it intractable if we require a numerical solution.
However, using the discrete approach, the addition of an extra parameter
or feature adds, at worst, a constant time factor to the time required
by the algorithm. The biological realism is central, as it may well be that case that
a novel component of the system that has not previously been considered proves to
be significant to the success or failure of the model. Traditional approaches generate biologically feasible
patterns, but we seek a degree of resolution that may not be achievable
without using a more fine-grained model. In addition, it is not clear
that these models are capable of generating regular patterns of both
``rings" and ``dots", which may be required by the anticipated
applications of the patterns (see Section ~\ref{applications}). Differential equation models do not treat each cell or subset of cells as individuals behaving in
a stochastic fashion. Our proposed approach uses a discrete approach that
models the behaviour of a relatively fine-grained subset of cells
which will yield the group-level pattern that might emerge from an
actual experiment. The degree of granularity of the simulation can
be easily modified by altering the number of bacteria represented
by each walker. In addition, bacteria such as \textit{ E. coli }move
at a constant rate, so it is easy to ``clock" the simulation and
ensure global synchronisation. By making the simulation spatially
 explicit and discrete (and therefore relatively easy to parallelise),
 we can easily take advantage of the computational power offered by
 cluster machines and the grid.

Support for this {\it cell-centred} view of biological modelling has gathered
momentum in the past few years. As Merks and Glazier argue in a recent article \cite{merks05}, ``...in many aspects of biological development, cells' inner workings are
irrelevant: what matters are the cell's biophysical properties, the signals it emits and its response to extracellular signals." While we would caution that their argument is from a
{\it developmental} perspective, and that we may not be able to fully treat the cell as a ``black box" for our purposes, we believe that the fundamental message is sound -- that cell-centred models will play an increasingly significant role in both generating hypotheses about and explaining collective cellular behaviours.
\newpage
\subsection{Fundamental issues}

Some of the fundamental questions that we seek to
address in the future  are: {\bf (1)} What level of complexity at the individual (bacterial) level is required to generate the observed complexity at the collective level? Additionally, how much of the observed complexity at the collective level
is a reflection of environmental rather than individual complexity?
{\bf (2)} Is it possible to simulate the growth and movement of bacterial
populations such that the complex patterns they generate {\it in
vitro} can be {\it specified in advance} in computationally
feasible times? {\bf (3)} If achieved, how may the associated abstract computational architectures, as well as the directed self-assembly of bacteria \textit{ in vitro} be applied
in various domains?
The first question is of fundamental importance to researchers in the field of biocomplexity.
The second and third questions are important if the great commercial potential
for this work can be realised. We propose using the model not only to further our understanding of collective cellular behaviour and chemotaxis, but as a framework for  {\it directed} pattern formation.  The fundamental idea is to develop a biologically valid simulation
of bacterial self-organisation, \textit{via continual and rigorous
testing against biological reality}. This laboratory work is {\it absolutely
crucial} to to the development of a reasonable model. Without this feedback from the biologists, the model will be nothing more than a very crude approximation of reality.
We may then allow a user to
specify a desired final pattern of bacteria, which may be useful
for a particular application such such as bio-sensing. Creating this pattern by
top-down placement of individual bacteria is possible \cite{xu}, but may be
unfeasible in the environments of practical interest,
so we seek to set up the system such that the pattern
self-organizes from an initial configuration or parameters (cell
density, initial placement of nutrients, gel structure, etc.). The
parameter space in which suitable configurations are located is
huge and multi-dimensional, so we may use an evolutionary
algorithm (EA) metaheuristic to search it more efficiently. EAs
are well-known for highly effective performance in large
combinatorial, numeric and mixed combinatorial/numeric spaces (as
in the case of the configuration space here), and recent work has already showed their efficacy in searching a space of {\it in vitro} bacterial colony
compositions \cite{van}. We could use the EA to
negotiate the space of initial configurations and parameter
settings, and the fitness function (which will be multiobjective) will simulate the model and
then calculate a vector of similarity measures between the desired
and final pattern. This forward simulation approach is vital,
because the desired pattern formation can take several days in the
laboratory. By running several million simulations in this time,
each with different parameter settings, we can efficiently
identify appropriate values for parameters that interact in
complex non-linear ways and minimise the amount of experimental
laboratory work required. We now consider possible practical implications of the extended model that we have just described.

\section{Possible applications}
\label{applications}

Models of chemotaxis have already been successfully applied
to the training of artificial neural networks \cite{brem,delgado}, multimodal function optimisation \cite{muller}, the design of aerofoils \cite{muller}, the control of robots for environmental monitoring \cite{dhar}, and the modelling of pedestrian evacuation processes \cite{kirch}. It is clear that this phenomenon has inspired a powerful optimisation strategy that is competitive with existing nature-inspired methods. However, what is lacking from the majority of these chemotactic models is a consideration of how the highly-localised interactions of the bacteria generate long-range biologically-advantageous patterns. By including such mechanisms, we may further enrich the possibilities for {\it distributed} bacterial optimisation. 

Perhaps more significantly, the physical patterns of bacteria will find
many applications in the biological and engineering domains. Self-assembly will
clearly play a strong role in determining the future direction of
(bio) nanotechnology \cite{white}. Our model would have implications for any application
that requires spatially registered cellular engineering. These may include the
development of novel biomaterials, the design of biosensors, and tissue engineering \cite{Amos04}.
We also anticipate its eventual application to the construction of structured microbial
consortia \cite{alp}, whose functioning in applications such as wastewater treatment and bioremediation is dependent on their precise structural ordering. Recent work has achieved small-scale, local directed pattern formation, using genetically-modified bacteria \cite{basu05}. By incorporating genetic factors with bacterial movement, our simulation framework would provide a powerful tool for the emerging field of {\it synthetic biology} \cite{sbeditorial,synthbio}

\section{Conclusions}

In this position paper we have outlined an ambitious vision of harnessing the complexity of bacterial communication and self-organisation for the purposes of human-defined computation and engineering. Rather than simply {\it observing} and modelling complex pattern formation, we seek to {\it control} and utilise it. Quite apart from the massive potential benefits to biology, we wish to understand a real and complex biological system so throughly that we can {\em control} it and harness its computational power to provide a wide range of potential {\it applications}. In doing so we expect to establish a highly
complex computational architecture for local-and-environmental-interaction
based pattern generation, from which we may extract several simplifications
and extensions with varieties of combinatorial properties. All of these
we expect to find applicable in many of the several high-impact sciences and
technologies emerging across the board which involve (or will benefit
from) the need for functional, adaptable and useful self-organisation of
collections of locally interacting and environmentally responsive units
(e.g., softbots in a network topology, self-organising and
responsive biosensors, biofilters, nanodevices, MEMs, smart skin
technologies, functional and programmable materials, cellular electronics,
environmental control, bioremediation). We now call for a concerted research effort in this direction.

\section*{Acknowledgements}

The authors are grateful to Jim Shapiro and David Corne for useful discussions.

\bibliography{./AmosHodgsonGibbons}

\end{document}